\documentclass[aps,prb,twocolumn,showpacs,superscriptaddress,groupedaddress]{revtex4}  % Uncomment for review/Submission version
\usepackage{graphicx}
\usepackage{sidecap}
\usepackage{dcolumn}
\usepackage{bm}
\usepackage{amssymb}
\usepackage{amsmath}
\usepackage{color}
\usepackage{multirow}
\begin{document}

%----------------------------------------------------------------------
% Title
%----------------------------------------------------------------------

\widetext
%\centerline{\textbf{Under preparation draft version - Please do not distribute}}
\title{Experimental and Atomistic Theoretical Study of Degree of Polarization \\ from Multi-layer InAs/GaAs Quantum Dot Stacks}

%----------------------------------------------------------------------
% Authors, affiliations and date
%----------------------------------------------------------------------

%\affiliation{Tyndall National Institute, Dyke Parade, Cork, Ireland}
%\affiliation{Physics Department, University College Cork, Cork, Ireland}

\author{Muhammad Usman} \email{usman@alumni.purdue.edu} \affiliation{Tyndall National Institute, Dyke Parade, Cork, Ireland} 
\author{Tomoya Inoue} \affiliation{Department of Electrical and Electronic Engineering, Graduate School of Engineering, 1-1 Rokkodai, Nada, Kobe 657-8501, Japan}
\author{Yukihiro Harda} \affiliation{Department of Electrical and Electronic Engineering, Graduate School of Engineering, 1-1 Rokkodai, Nada, Kobe 657-8501, Japan}
\author{Gerhard Klimeck} \affiliation{Purdue University, Network for Computational Nanotechnology, Birck Nanotechnology Center, 1205 W. State Street, West Lafayette, IN 47907, USA}
\author{Takashi Kita} \affiliation{Department of Electrical and Electronic Engineering, Graduate School of Engineering, 1-1 Rokkodai, Nada, Kobe 657-8501, Japan}
\vskip 0.25cm
%\date{\today}

%----------------------------------------------------------------------
% Abstract
%----------------------------------------------------------------------

\begin{abstract}
Recent experimental measurements, without any theoretical guidance, showed that isotropic polarization response can be achieved by increasing the number of QD layers in a QD stack. Here we analyse the polarization response of multi-layer quantum dot stacks containing up to nine quantum dot layers by linearly polarized PL measurements and by carrying out a systematic set of multi-million atom simulations. The atomistic modeling and simulations allow us to include correct symmetry properties in the calculations of the optical spectra: a factor critical to explain the experimental evidence. The values of the degree of polarization (DOP) calculated from our model follows the trends of the experimental data. We also present detailed physical insight by examining strain profiles, band edges diagrams and wave function plots. Multi-directional PL measurements and calculations of the DOP reveal a unique property of InAs quantum dot stacks that the TE response is anisotropic in the plane of the stacks. Therefore a single value of the DOP is not sufficient to fully characterize the polarization response. We explain this anisotropy of the TE-modes by orientation of hole wave functions along the [$\bar{1}10$] direction. Our results provide a new insight that isotropic polarization response measured in the experimental PL spectra is due to two factors: (i) TM$_{001}$-mode contributions increase due to enhanced intermixing of HH and LH bands, and (ii) TE$_{110}$-mode contributions reduce significantly due to hole wave function alignment along the [$\bar{1}10$] direction. We also present optical spectra for various geometry configurations of quantum dot stacks to provide a guide to experimentalists for the design of multi-layer QD stacks for optical devices. Our results predict that the QD stacks with identical layers will exhibit lower values of the DOP than the stacks with non-identical layers.     
\end{abstract}

\maketitle

%----------------------------------------------------------------------
% Main Text
%----------------------------------------------------------------------

\textbf{\textit{Introduction and Problem Background:}} The design of optical devices such as semiconductor optical amplifiers requires polarization insensitive optical emissions. The polarization response of quantum dots is measured in terms of degree of polarization (DOP), defined as:

\begin{equation}
DOP = \frac{TE_{\bot -growth}-TM_{\parallel -growth}}{TE_{\bot -growth}+TM_{\parallel -growth}} 
\end{equation} 

Here TE$_{\bot -growth}$ refers to traverse electric mode in a direction perpendicular to the growth direction ([001]) and TM$_{\parallel -growth}$ refers to traverse magnetic mode along the growth direction ([001]). The value of DOP depends on the chosen direction for the TE-mode in the plane of quantum dot. While the previous studies of polarization response of quantum dots provide only one value of DOP corresponding to a chosen direction for the TE-mode\cite{Ridha_1,Ridha_2,Inoue_1,Saito_1}, we associate DOP with the direction of TE-mode, for example TE$_{110}$ $\rightarrow$ DOP$_{110}$ and TE$_{\bar{1}10}$ $\rightarrow$ DOP$_{\bar{1}10}$, and show that the DOP depends highly on the chosen direction. Only one DOP value is not sufficient to fully characterize the polarization response of quantum dots systems. 

The InAs quantum dots obtained from the Stranski-Krastanov self-assembly growth process typically have a flat shape i.e. base diameter is typically 4-5 times larger than the height. In such quantum dots, compressive biaxial strain splits the heavy hole HH and light hole LH bands by more than 100meV. As a result, only the TE-mode can couple and the TM-mode is very weak\cite{Singh_1}. The polarization response of such systems is highly anisotropic, TE-mode $\gg$ TM-mode and DOP$\rightarrow$1.0. To achieve the desired isotropic response (DOP=0) for the design of optical devices, significant tuning of QD geometry, band structure manipulation, and/or strain engineering are required.

\begin{figure*}
\includegraphics[scale=0.42]{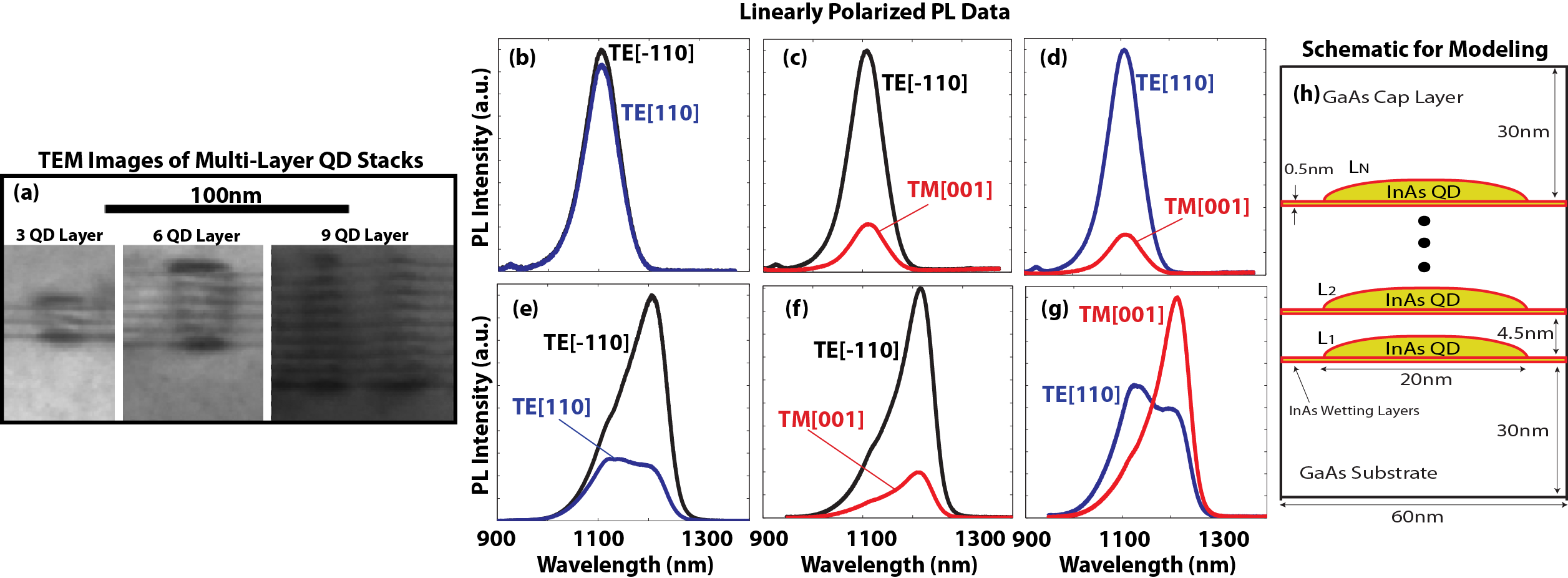}
\caption{(a)TEM images of three, six, and nine QD layer stacks grown by solid-source molecular-beam epitaxy\cite{Inoue_1}. The layers of InAs QDs separated by GaAs intermediate layers are stacked with clear wetting layer interfaces. (b, c, d) Linearly polarized PL measurements for TE$_{110}$, TE$_{\bar{1}10}$, and TM$_{001}$-modes for single QD layer. (e, f, g) Linearly polarized PL measurements for TE$_{110}$-, TE$_{\bar{1}10}$-, and TM$_{001}$-modes for nine QD layers sample. The PL plots in the figures (b) to (g) are from independent experimental measurements. Each one of them is normalized to its highest peak. (h) The schematic diagram of the simulated system. InAs quantum dot layers are embedded inside GaAs buffer. Each layer consists of a dome-shaped quantum dot on top of 1.0ML InAs wetting layer. QD layers are denoted by L$_N$, where N being the number of QD layers in the stack. The wetting layers are separated by 4.5nm GaAs buffer. }
\label{fig:system}
\end{figure*}
\vspace{1mm}   
 
During the last few years, several techniques have been explored to achieve polarization insensitive optical emission from InAs quantum dot (QD) samples. These methods include overgrowing the InAs QD samples by an InGaAs strain relaxing capping layer (SRCL)\cite{Usman_1, Tatebayashi_1}, growing large stacks of QDs in the form of columnar QDs\cite{Ridha_1, Saito_1, Kita_1, Kita_2}, bi-layer \cite{Usman_2}, tri-layer QD\cite{Fortunato_1}, multi-layer stacks\cite{Kojima_1, Kojima_2, Inoue_1}, and band gap engineering by including dilute nitrogen N\cite{Richter_1}, phosphorous P\cite{Jiang_1}, and antimony Sb\cite{Liu_1} impurities. T. Kita \textit{et al.}\cite{Kita_2} and T. Saito \textit{et al}.\cite{Saito_1} demonstrated that an isotropic polarization response can be obtained by growing columnar QDs consisting of nine QD layers. 

Recent experiments by T. Inoue \textit{et al.}\cite{Inoue_1,Inoue_2,Inoue_4} showed, without any theoretical guidance, that similar tuning of polarization properties is possible in regular InAs QD stacks where the QD layers are geometrically separated by thin GaAs spacers. Such multi-layer QD stacks have a twofold advantage over columnar QDs: (i) a moderately thick GaAs spacer between the QD layers allows a precise control of overall QD shape and size, (ii) a reduced strain accumulation results in isotropic polarization response with fewer number of QD layers in the stack. The experimental results are provided for QD stacks containing three, six, and nine QD layers. The results indicate that the DOP$_{110}$ takes up the values of +0.46 and -0.60 for the samples containing six and nine QD layers, respectively. A change of sign for the DOP$_{110}$ implies that an isotropic polarization response (DOP$\sim$0) can be achieved by engineering the number of QD layers in the stack.

The previous experimental measurements\cite{Inoue_1} indicated a prospect to achieve polarization insensitive response, however no theoretical study is available to-date to provide physical insight for the design of these complex multi-million atom nano-structures. Furthermore the experiments only analyse TE$_{110}$-mode whereas PL measurements along multiple in-plane directions are required to fully characterize the polarization properties of these QD systems. 

This article aims to thoroughly investigate the optical spectra of multi-layer QD stacks by in-plane polarization measurements and atomistic theoretical analysis. Our PL measurements reveal unique properties of these QD stacks containing strongly coupled electronic states and the atomistic theory explains the experimental evidence very well. The modeling and simulations also provide data for DOP for different geometry configurations and in-plane directions for the TE-mode to explore the design space and provide a guide for future experiments.  

\textbf{\textit{Experimental Procedure:}} The QD samples are grown on an undoped [001] GaAs subtrate using solid-source molecular-beam epitaxy. First a thick layer of GaAs buffer layer is grown at $550^\circ$C, followed by repetitions of the InAs QD layers and GaAs spacer layers with growth interruptions of 10s after each GaAs spacer layer reached a thickness of 16MLs. The nominal thickness of the InAs is kept 1.9ML. Finally the QD layers are capped with a 100nm thick GaAs layer. Further details of our growth process can be found in earlier publications\cite{Inoue_1, Inoue_2, Inoue_3, Inoue_4}. The crystallographic properties of the stacked QDs are examined using a cross-sectional transmission electron microscope (TEM). The TEM images of the samples containing three, six, and nine QD layers are shown in Fig. ~\ref{fig:system}(a)\cite{Inoue_1}. 

Next, we perform linearly polarized PL measurements at room temperature to investigate the polarization dependent optical spectra. The laser diode excitation is at 659nm wavelength. The detailed set up and measurement procedure is described by T. Inoue \textit{et al}\cite{Inoue_2}. The polarization dependent PL spectra are shown in Fig. ~\ref{fig:system} for a single QD layer (b, c, d) and nine QD layers (e, f, g).

Fig. ~\ref{fig:system}(b, c, d) indicate for a single quantum dot that TE$_{110}$ $\sim$ and TE$_{\bar{1}10}$ show a similar magnitude, and that TM$_{001}$ has a much smaller response than both in-plane TE-modes. This is typical for InAs QDs and is due to the compressive biaxial strain that splits the HH-LH band edges resulting in HH-type valence band states close to band gap. Previous theoretical and experimental studies on single QDs have shown similar properties.

The PL spectra in the Fig. ~\ref{fig:system}(e, f, g) on contrary reveal an interesting and unique property of the InAs/GaAs QD stacks that the TE$_{110}$ response is significantly less than the TE$_{\bar{1}10}$ response. No previous evidence exists for such TE-mode anisotropy in InAs/GaAs QDs, though Podemski \textit{et al.}\cite{Podemski_1} have reported similar results for InP-based columnar quantum dash structures. The measured difference between the two TE-modes is such that even the DOP for the same QD stack could have different signs when measured along the [110] and [$\bar{1}10$] directions. The reason for such TE-mode anisotropy is highly non-intuitive and requires modeling and simulations of these large QD structures. The modeling must include the correct symmetry properties because the long range strain effects the room temperature material properties and atomistic structure resolution. The multi-million atom simulations presented in this article explains the experimental measurements in terms of hole wave function alignments along the [$\bar{1}$10] direction.              

\textbf{\textit{Theoretical Model:}} The theoretical modeling of QD stacks containing up to nine QD layers posses a twofold challenge: first it requires an atomistic model that can calculate electronic and optical properties including correct symmetry and interfaces. Secondly the large size of QD stacks requires calculations to be done over millions of atoms to properly include the long-range effects of strain. We use NEMO 3-D\cite{Klimeck_1, Klimeck_2} to analyse the electronic and optical properties of multi-layer QD stacks. NEMO 3-D is based on fully atomistic calculations of strain and electronic structure: the strain is calculated by using atomistic valence force field (VFF) model\cite{Keating_1}, including anharmonic corrections\cite{Lazarenkova_1} to the Keating potential and the electronic structure is calculated by solving a Hamiltonian inside twenty band $sp^{3}d^{5}s^{*}$ basis\cite{Boykin_1}. 

NEMO 3-D has been designed and optimized to be scalable from single CPU to large number of processors on most advanced supercomputing clusters. Excellent MPI based scaling to 8192 cores/CPU has been demonstrated\cite{Ahmed_1}. The atomistic modeling techniques and parallel coding scheme implemented in NEMO 3-D allows it to simulate large QD stacks with realistic geometry extension and symmetry properties. Past NEMO 3-D based studies of nano-structures include (i) single InAs QD with InGaAs SRCL\cite{Usman_1}, InAs bilayer QDs\cite{Usman_2,Usman_4}, and InAs multi-layer QD stacks\cite{Marek_1}, (ii) valley splitting in miscut Si quantum wells on SiGe substrate\cite{Kharche_1}, (iii) Stark effect of single P impurities in Si\cite{Rahman_1}, and (iv) dilute Bi impurities in GaAs and GaP materials\cite{Usman_5}.     
  
\textbf{\textit{Simulated System:}} The theoretical analysis of the multi-layer QD stacks containing up to nine layers is carried out through a set of systematic simulations. The geometry parameters of the QD samples are extracted from the TEM images shown in the Fig. \ref{fig:system}(a) which indicates that all the QDs in the stacks are of nearly identical size. Some top most layers in the nine quantum dot stack appears to be relatively small and are only partially grown\cite{Inoue_1}. The schematic diagram of the system modelled in our simulations is shown in Fig. ~\ref{fig:system}(h). Multiple QD layers separated by 4.5nm thick GaAs buffer are embedded inside a large GaAs matrix. This wetting layer-to-layer separation is experimentally optimized to obtain uniform vertical QD stacks from the self-assembly growth process\cite{Inoue_2}. Each QD layer consists of a dome-shaped InAs QD with circular base, lying on top of a 1.0ML InAs wetting layer. The QD systems with single, three, six, and nine QD layers will be labelled as L$_1$, L$_3$, L$_6$, and L$_9$, respectively. 

The size of QDs extracted from the TEM images indicates a base diameter of $\sim$20nm and height of $\sim$4nm. Since the height of the quantum dots in the TEM images is not very clear, we choose to simulate three different QD geometries in our theoretical study: (i) All QD layers are identical with 20nm base diameter and 4nm height of the QD in each layer. All the results presented are for this system unless other dimensions are specified. (ii) All QD layers are identical with 20nm base diameter and 3.5nm height of the QD in each layer and (iii) the size of QDs increases from the lower to the upper layers: the base diameter increases by 1nm and the height increases by 0.25nm. We choose this last system (iii) because past experiments of multi-layer QD stacks\cite{Xie_1, Usman_2} have indicated an increasing size of QDs when multi-layer QD stacks are grown by the self-assembly process, so it is interesting to theoretical investigate this geometry. We also simulate a system L$_9$ in which the height of QDs in each layer is 4.5nm. This would mean that the top of each quantum dot will be touching its upper adjacent wetting layer, approaching columnar QD limit.  The theoretical results presented here for various QD geometries provide a guide for the experimentalists to understand the dependence of DOP on the QD geometry since experimental investigations of QD stacks for polarization response are still under way.     

The QD layers are embedded inside a sufficiently large GaAs buffer to ensure proper relaxation of atoms and to accommodate the long-range effects of strain. The size of the largest GaAs buffer for the system containing nine QD layers L$_9$ is 60x60x106$nm^3$, consisting of $\sim$24.4 million atoms. Mixed boundary conditions are applied in the strain minimization: the substrate is fixed at the bottom, the GaAs matrix is periodic in lateral dimensions, and the capping layer is free to relax from the top. The electronic structure calculations use a separate subdomain with closed boundary conditions to reduce the computational burden~\cite{Klimeck_2}, with a surface passivation~\cite{Lee_1} which avoids artificial surface states in the atomistic representation.

\begin{figure}
\includegraphics[scale=0.28]{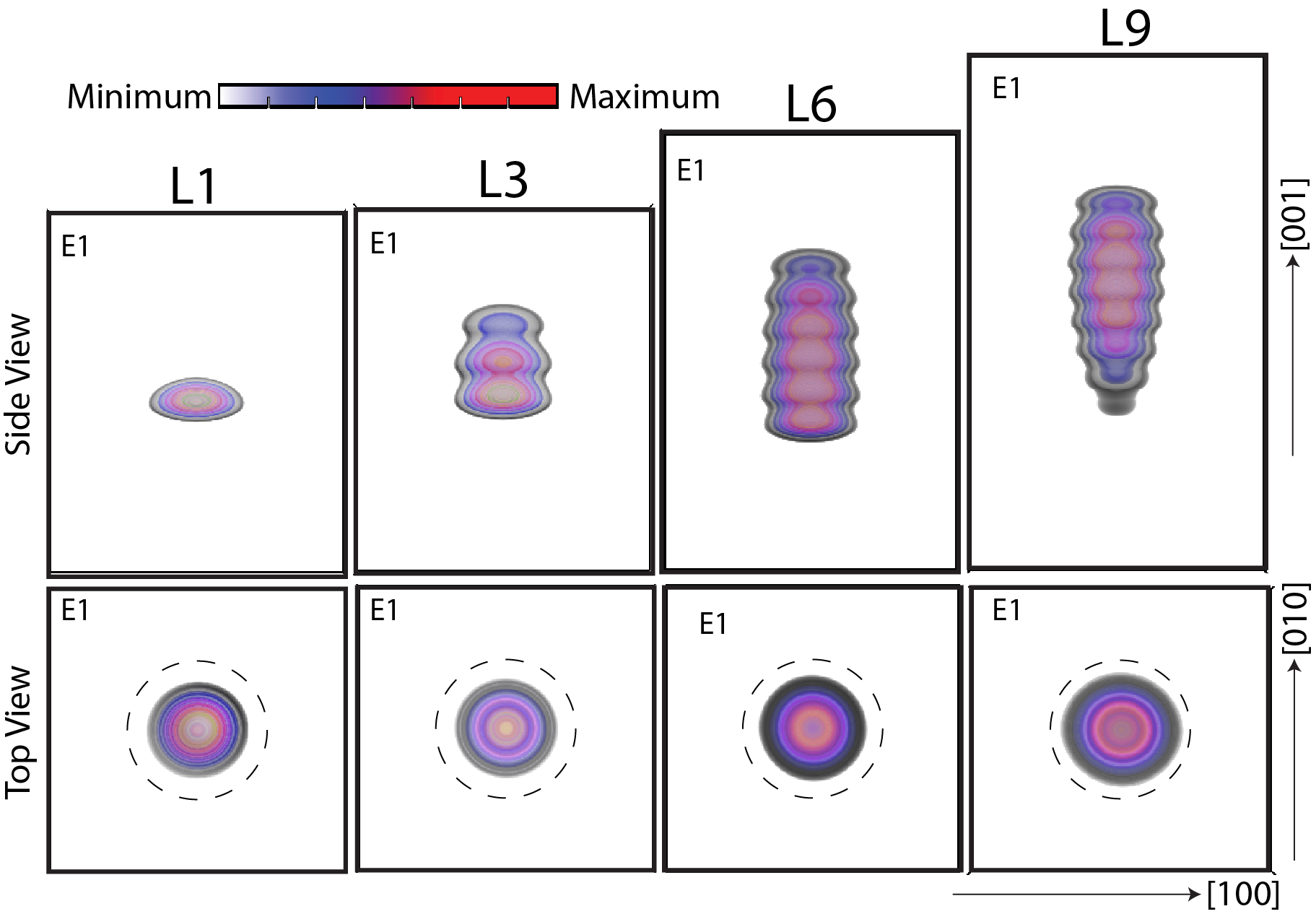}
\caption{Plots of the lowest conduction band state, E$_1$, for quantum dot systems L$_1$, L$_3$, L$_6$, and L$_9$. First row: the side view of the plots is shown. Second Row: the top view of the plots is shown. The intensity of the color in the plots indicates the magnitude of the wave function: the red color represents the highest magnitude and the light blue color represents the lowest magnitude. The dotted circles are marked to guide the eye and indicate the boundary of the base of each QD. }
\label{fig:electron_wf}
\end{figure}

\textbf{\textit{Electron Wave functions form molecular states:}} Figure ~\ref{fig:electron_wf} shows the plots of the lowest conduction band state E$_{1}$ for all of the four QD systems under study. From the top views of the wave functions (second row), it is evident that the lowest electron state is of s-type symmetry. The side views of the wave functions (first row) show that the electron state forms a hybridized (molecular) state in L$_{3}$, L$_{6}$, and L$_{9}$ stacks, and is spread over all of the quantum dots. This is due to strong coupling of quantum dots at 4.5nm separation. The presence of the s-like electron wave function in all of the quantum dots implies that only the details of hole wave functions inside the quantum dot stack will determine the optical activity of a particular quantum dot inside the stack. This is different from the previous study of bilayers \cite{Usman_2}, where a weak coupling of quantum dots at 10nm separation resulted in atomic like electron wave functions. In the bilayer system, both the electron and the hole wave functions determined the optical activity of the quantum dots. Previous studies on the identical bilayer\cite{Usman_4} and stacks with seven identical QDs\cite{Marek_1} showed that the strain tends to push the electron states towards the lower QDs in such systems. The electron wave functions for L$_3$ and L$_6$ systems follow this trend. Here we find that this trend is no longer true for stacks with nine QD layers, L$_9$, where the electron wave function E$_1$ vanishes around the edges of the stack due to the larger strain magnitude there (see Fig. ~\ref{fig:electron_wf} first row for L$_9$).  

\vspace{1mm}

\begin{figure}[b]
\centering
\includegraphics[width=0.5\textwidth]{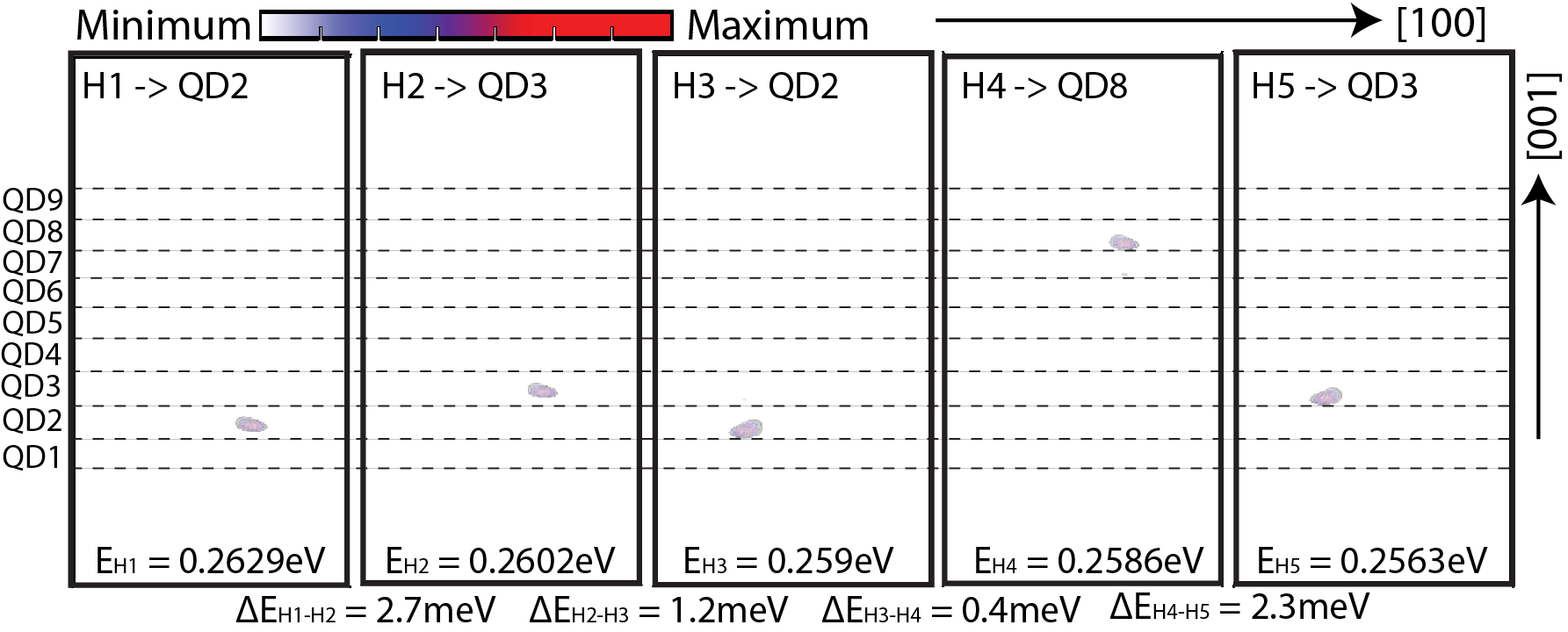}
\caption{Plots of the highest five valence band state, H$_1$, H$_2$, H$_3$, H$_4$ and H$_5$, for quantum dot systems L$_9$. Only the side view of the plots is shown. The horizontal dotted lines are marked to guide the eye and indicate the base of the QD layers in the stack. The intensity of the color in the plots indicates the magnitude of the wave function: the red color represents the highest magnitude and the light blue color represents the lowest magnitude. The energies of the valence band states and the differences between the energies of the adjacent levels are also mentioned.}
\label{fig:hole_side_view}
\end{figure} 

\textbf{\textit{Hole wave functions exhibit atomic character:}} Although the electrons (lighter-mass particles) are strongly influenced by the inter-dot electronic and strain couplings of QDs and exhibit tunnelling across the quantum dots forming molecular like hybridized states, the holes due to their heavier mass remain well confined inside the individual dots and do not show any hybridization \cite{Bester_1}. For example, Fig. ~\ref{fig:hole_side_view} show the side views of the lowest five valence band wave functions in the L$_9$ system. The horizontal dotted lines are plotted to mark the positions of the base of the QDs and helps to determine the location of a particular hole wave function inside the stack. In this system, H$_1$ and H$_3$ are inside QD$_2$, H$_2$ and H$_5$ are in QD$_3$, and H$_4$ is in QD$_8$. The location of a hole state inside a QD stack is relatively hard to determine and is strongly influenced by geometry of the QD stack \textit{i.e.} QD base diameter, QD height, QD layer separation etc. Ultimately the strain profile that controls the strength of the coupling between the QD layers inside the stack determines the position of the hole states inside the stacks.         

\begin{figure}
\includegraphics[scale=0.45]{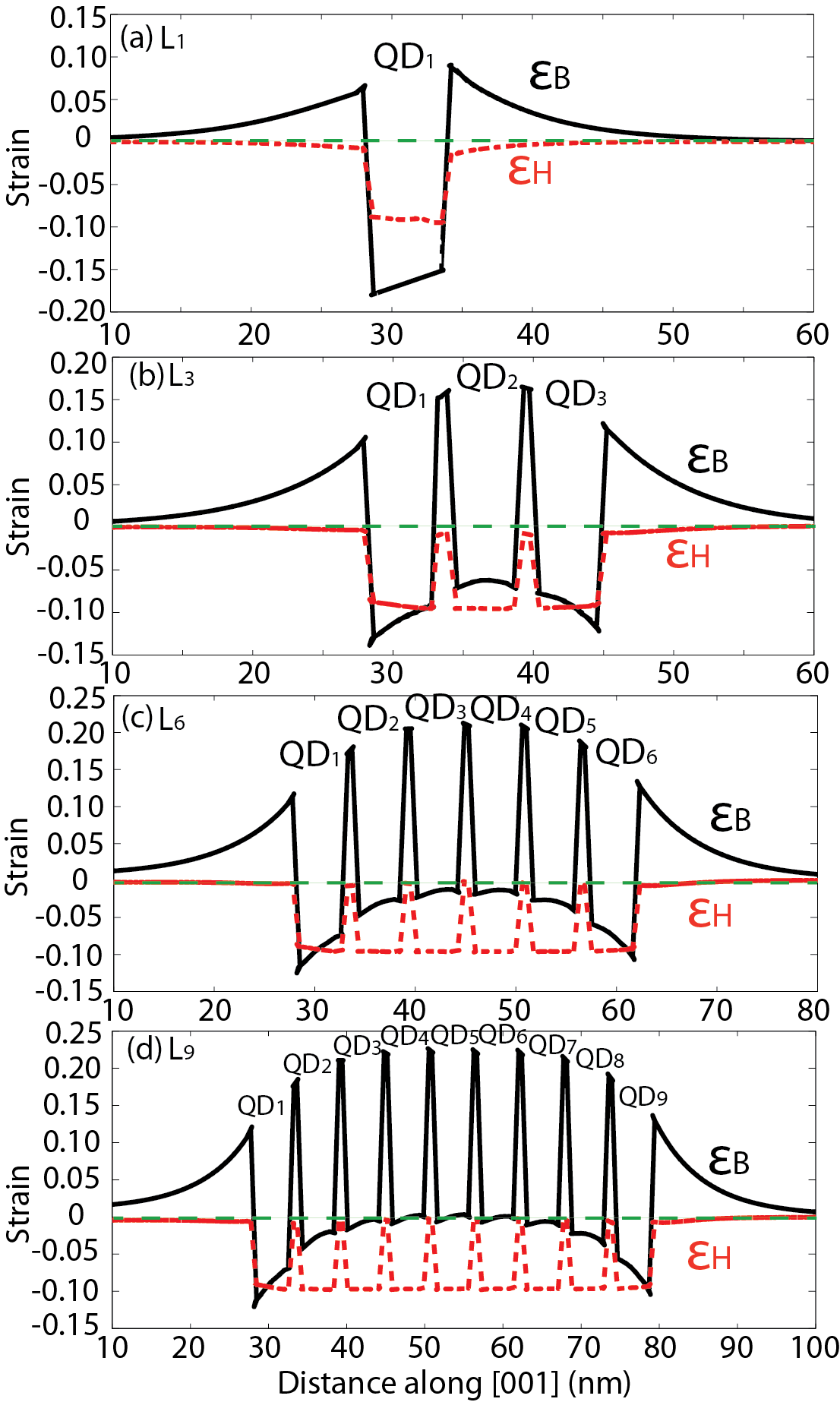}
\caption{The plots of hydrostatic ($\in_{H}=\in_{xx}+\in_{yy}+\in_{zz}$) and biaxial strain ($\in_{B}=\in_{xx}+\in_{yy}-2\in_{zz}$) components through the center of the quantum dot system along the [001] direction. The hydrostatic component is dominantly negative inside the QD indicating strong compression of the InAs and alomost zero outside the QD. As the QD stack height increases, the biaxial strain in the QD evolves from negative to zero with a small increase in positive contributions in the capping layer.}
\label{fig:strain}
\end{figure}

\textbf{\textit{Hydrostatic and Biaxial Strains:}} Figure ~\ref{fig:strain} plots the hydrostatic $\in_{H}=\in_{xx}+\in_{yy}+\in_{zz}$ (dotted lines) and biaxial strain $\in_{B}=\in_{xx}+\in_{yy}-2\in_{zz}$ (solid lines) profiles along the [001] direction through the center of the quantum dots. The hydrostatic strain exhibit a very slight change from L$_1$ to L$_9$. The biaxial strain however significantly changes as the QD stack height increases. For a single QD, L$_1$, the biaxial strain is highly negative inside the QD region. As the vertical size of the QD stack is increased by adding QD layers, the biaxial strain at the center of the stack reduces. For the L$_9$ stack in the Fig. ~\ref{fig:strain}(d), the biaxial strain at the center of the stack approaches zero. The reason for such behaviour of the biaxial strain is that in general the InAs unit cells inside the QD region tend to fit over the GaAs matrix by an in-plane compression and an elongation along the [001] direction. This results in highly negative biaxial strain as can be observed for a single QD in the Fig.~\ref{fig:strain}(a). However, when the size of the stack increases, the unit cells of InAs around the center of the stack feel lesser and lesser compressive force from the surrounding GaAs. As a result, the vertical lattice constant of the InAs starts matching with the GaAs and hence the biaxial strain tends to change its sign around the middle of the QD stack. Similar strain profiles were calculated in an earlier study about columnar QDs by T. Saito \textit{et al.}\cite{Saito_1}.

\vspace{3mm} 

\begin{figure}
\includegraphics[scale=0.56]{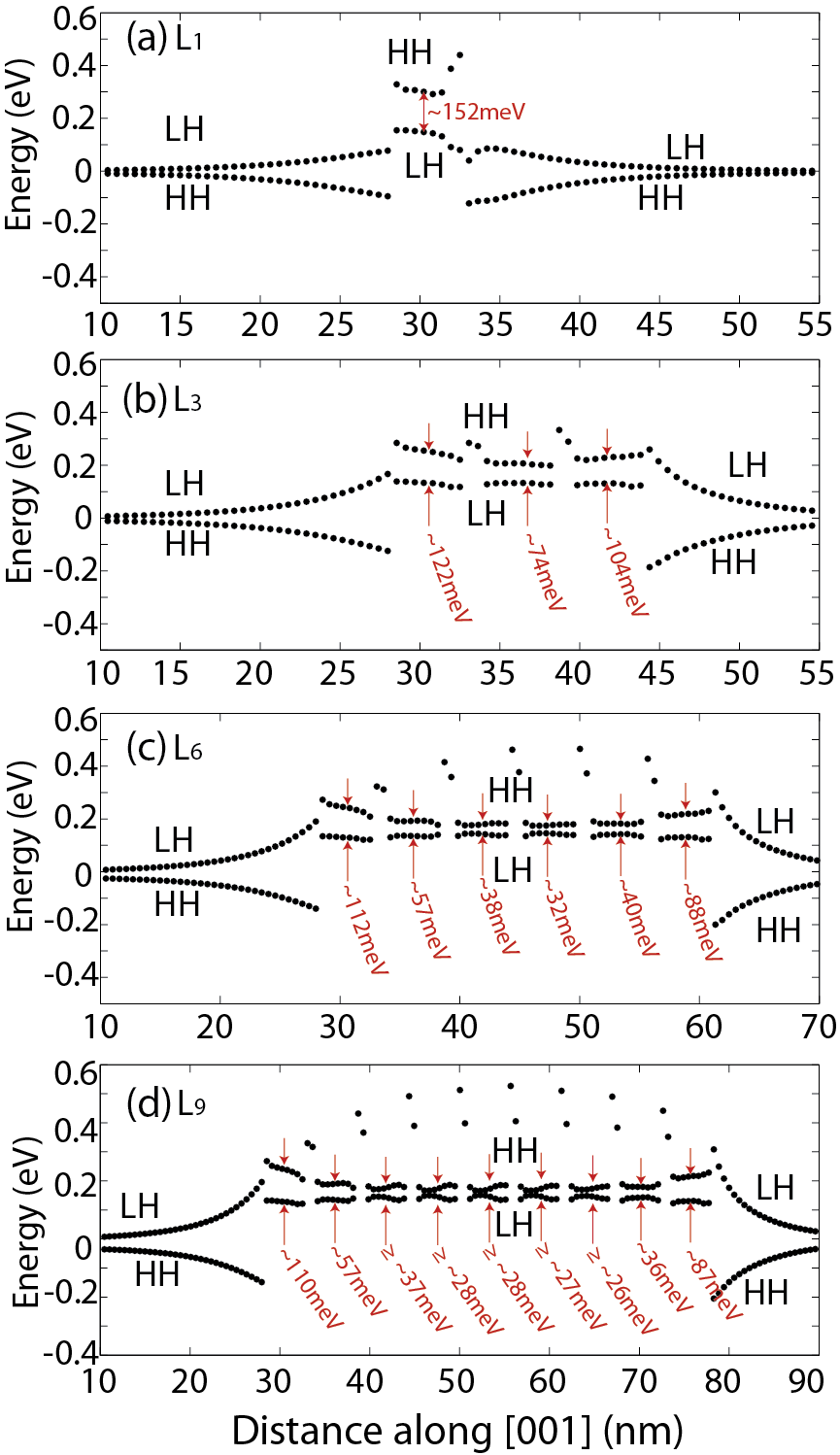}
\caption{Plots of local band edges for the highest two valence bands (HH and LH) through the center of the QD along the [001] direction for L$_1$, L$_3$, L$_6$, and L$_9$. The reduction in the magnitude of biaxial strain results in larger HH/LH intermixing (lesser separation) as the size of the stack increases.}
\label{fig:local_bands}
\end{figure}
\vspace{3mm}   

\textbf{\textit{Increased HH-LH Mixing:}} The minor change in the magnitude of the hydrostatic strain (as L$_1$ $\rightarrow$ L$_9$) implies that the lowest conduction band edge will experience very small change as they are only effected by the hydrostatic component\cite{Usman_1}. The valence band edges are affected by both the hydrostatic strain as well as the biaxial strain. The impact of strain on the highest two valence band edges, HH and LH, is analytically expressed as: 

\begin{equation}
\delta E_{HH} = a_{v}\in_{H} + \frac{b_{v}\in_{B}}{2}
\end{equation}
\begin{equation}
\delta E_{LH} = a_{v}\in_{H} - \frac{b_{v}\in_{B}}{2}  
\end{equation}

Here $a_v$ and $b_v$ are deformation potential constants for HH and LH band edges. The values for these constants for InAs systems are $a_v$ = 1.0eV and $b_v$ = -1.8eV, respectively\cite{Usman_1}. From the equations 2 and 3, it is evident that the magnitude of the $\in_{B}$ determines the HH-LH splitting. For a single QD layer, due to large negative value of $\in_{B}$, the HH and LH band edges will be considerably separated inside the QD region. This will induce dominant HH character in the highest few valence band states which will be closer to the HH band edge. As the magnitude of $\in_{B}$ decreases, the HH-LH splitting reduces, increasing LH component in the valence band states. For the L$_9$ system, the nearly zero magnitude of the $\in_{B}$ around the center of the stack implies that the HH and LH bands will be nearly degenerate around the center of the QD stack. The valence band states will therefore be of highly mixed character, consisting of contributions from both the HH and the LH bands. 

Figure ~\ref{fig:local_bands}(a-d) plots the highest two local valence band edges, HH and LH, for all of the QD systems under study along the [001] direction through the center of the QDs. Highly negative biaxial strain in L$_1$ results in $\sim$152meV splitting of HH and LH bands within the QD region. As the biaxial strain around the center of QD stack decreases (approaching towards zero), the HH-LH splitting around the center of the stack also decreases to $\sim$74meV, $\sim$32meV, and $\leq$28meV for the L$_3$, L$_6$, and L$_9$ systems, respectively. 

The HH and LH character of a particular valence band state in the tight binding formulation can be estimated as follows: If the amplitudes of the p$_{x}$, p$_{y}$, and p$_{z}$ orbitals at any atomic site are a$_{x, u/d}$, a$_{y, u/d}$, and a$_{z, u/d}$  respectively (where the subscripts $u$ and $d$ refer to up and down spin), then  the HH contribution is approximately proportional to $|a_{x,u}-ia_{y,u}|^2$+$|a_{x,d}+ia_{y,d}|^2$ summed over all the atomic sites. The LH contribution is approximately proportional to $|a_{z,u}|^2$+$|a_{z,d}|^2$ summed over all the atomic sites. By using these expressions, we estimate that the $\frac{HH}{LH}$ ratio for the highest valence band state (H$_1$) decreases from $\sim$108 for the L$_1$ to $\sim$15.8, $\sim$12.3, and $\sim$10.6 for the L$_3$, L$_6$, and L$_9$ systems, respectively. This clearly points towards an increasing LH character in the valence band states as the height of the QD stack increases.

\begin{SCfigure*}
\centering
\includegraphics[width=0.6\textwidth]%
{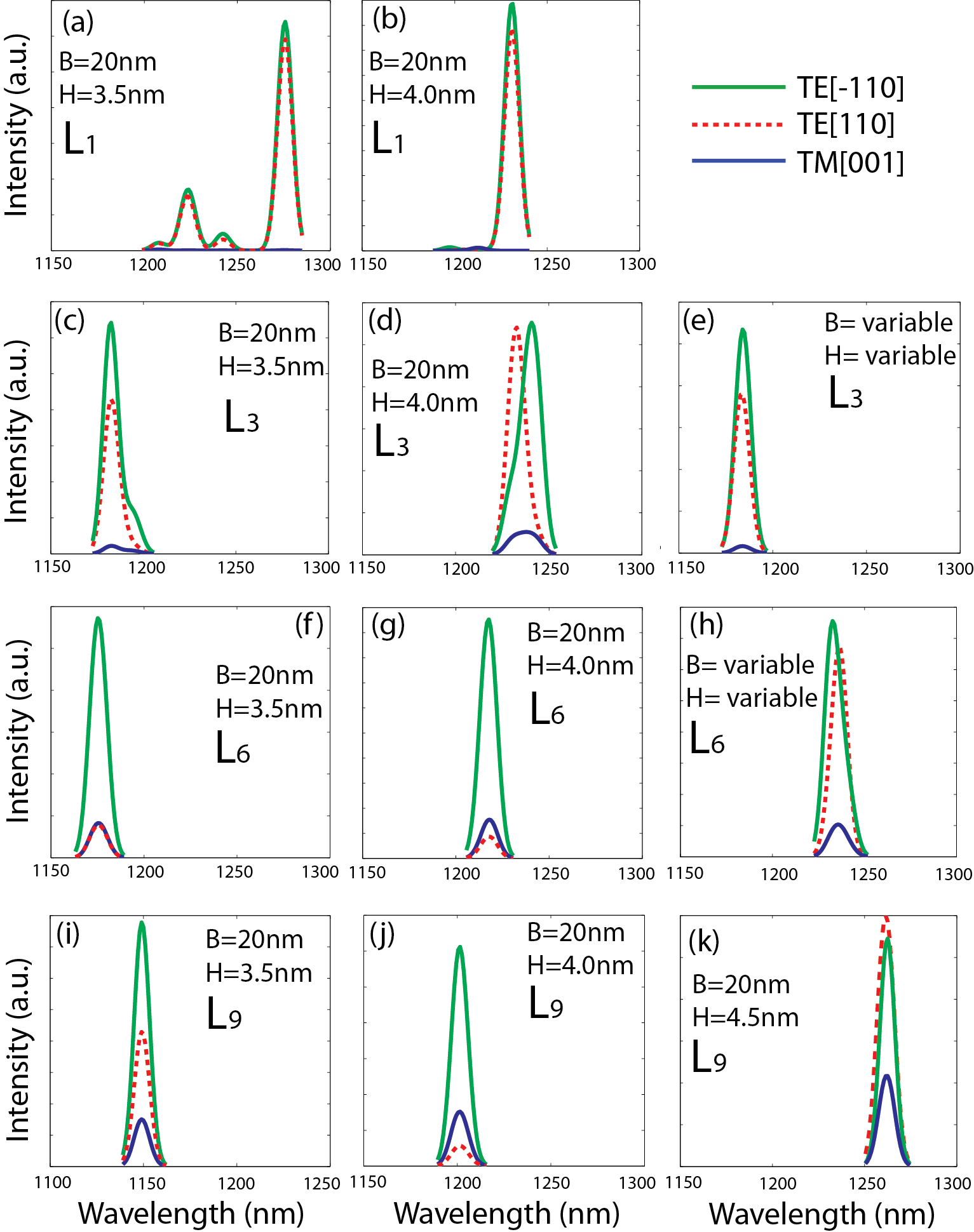}% picture filename
\caption{ The plots of optical intensity functions, $f(\lambda)$, are shown for various QD systems. The optical intensity functions in each case are computed from equations 4 and 5. The rows of the figure represent the QD system: first row=L$_1$, second row=$L_3$, third row=$L_6$, and fourth row=L$_9$. The columns of the figure represent the geometry of the particular system. First and second columns show results for identical QD stacks with 3.5nm and 4.0nm QD heights, receptively, and 20nm base diameter. The third column is a special case where we consider increasing size of QDs for L$_3$ and L$_6$ systems. For L$_9$, we again simulate identical QDs, but each with height 4.5nm. This is the case when regular QD stack approaches columnar QD shape.}
\label{fig:intensities}
\end{SCfigure*}
\vspace{2mm}

\textbf{\textit{HH/LH intermixing implies TM$_{001}$-mode increases:}} In a QD system, the HH states consist of contributions from p$_{x}$ and p$_{y}$ orbitals and the LH states consist of contributions from p$_{x}$, p$_{y}$, and p$_{z}$ orbitals\cite{Singh_1}. These configurations imply that the TM-mode (which is along the z-direction) will only couple to the LH states. The large splitting of the HH-LH bands (see Fig. ~\ref{fig:local_bands}(a)) resulting in a weak LH contribution in the L$_1$ system will result in very weak TM$_{001}$-mode for this system. Thus from Eq. (1), the DOP will be nearly 1.0 and the polarization response will be highly anisotropic. As the size of the QD stack increases, the larger intermixing of HH and LH bands increases LH contribution in the valence band states. This will result in increase of TM$_{001}$-mode of optical transitions reducing the anisotropy of DOP bringing it closer to 0.             

\textbf{\textit{Optical intensity functions, \textit{f($\lambda$)}:}} Figure ~\ref{fig:intensities} plots the optical intensity functions computed from our model as a function of the optical wavelength for various quantum dot systems. The calculation of the optical intensity function is done as follows: first we calculate optical transition strengths by using Fermi's golden rule\cite{Usman_2} for TE$_{110}$-, TE$_{\bar{1}10}$-, and TM$_{001}$-modes between the lowest conduction band state E$_1$ and the highest five valence band states H$_1$, H$_2$, H$_3$, H$_4$, and H$_5$ for a particular QD system. Next each optical transition strength is artificially broadened by multiplication with a Gaussian distribution centred at the wavelength of the transition\cite{Stier_1, Singh_2}. Finally we add all of the five Gaussian functions to calculate the total optical intensity function, \textit{f($\lambda$)}. The complete expression for the optical intensity function, plotted in the Fig. \ref{fig:intensities}(a-k), is given by equations 4 and 5:

\begin{equation}
{f(\lambda)}_{T^{E_1-H_i}} = \sum_{i=1}^{5} (T^{E_1-H_i}).e^{-\frac{\displaystyle \lambda - \lambda_{E_{1}-H_{i}}}{\displaystyle (0.25)^2}}
\end{equation}
where,
\begin{equation}
T^{E_1-H_i} = (TE_{110}^{E_1-H_i}/TE_{\bar{1}10}^{E_1-H_i}/TM_{001}^{E_1-H_i})
\end{equation}
\vspace{2mm}

The examination of the optical intensity plots in the Fig. ~\ref{fig:intensities} reveals that the TM$_{001}$-mode indeed increases as the size of QD stack is increased: L$_1$ $\rightarrow$ L$_9$. This is in general true for all of the geometry configurations considered and is a direct consequence of the change in the biaxial strain component that increases HH and LH intermixing as discussed earlier.

\textbf{\textit{Increase in TM$_{001}$ only partially contributes towards isotropic polarization:}} Fig. ~\ref{fig:intensities} shows that the increase in the TM$_{001}$-mode only partially helps towards an isotropic polarization response. This is in contrast to a general notion where it is described that the increase in the TM$_{001}$-mode is mainly responsible for the isotropic polarization.The reason for such understanding is that the previous theoretical\cite{Saito_1} or the experimental studies\cite{Inoue_1,Ridha_1,Fortunato_1} of the DOP have assumed only one direction for the TE-mode. However, our PL measurements in the Fig. ~\ref{fig:system}(e-g) show that the TE-modes along the [110] and [$\bar{1}10$] have significant anisotropy in the plane of the QD stack. The theoretical model shows that in fact a major contribution to achieve isotropic polarization response in these systems stems from a suppressed TE$_{110}$-mode rather than an increased TM$_{001}$-mode. Fig. ~\ref{fig:intensities} shows that irrespective of QD geometry, the increase in the TM$_{001}$-mode is insufficient to reverse the sign of the DOP$_{\bar{1}10}$ (+ $\rightarrow$ -).

The TE-mode is highly anisotropic in the plane of the QD with the magnitudes of the TE$_{110}$- and TE$_{\bar{1}10}$-modes becoming very different as the QD stack size increases. For a single QD system, L$_1$, TE$_{110} \sim$ TE$_{\bar{1}10}$ and TM$_{001}$ is very weak. Hence the measured and calculated DOP is highly anisotropic (close to 1.0) irrespective of the direction for the TE-mode. As the QD stack size increases, the TM$_{001}$-mode also increases partially contributing to reduction in the DOP. However, at the same time, the TE$_{110}$-mode decreases considerably such that for L$_6$ and L$_9$, it becomes smaller than the TM$_{001}$-mode. The reason for such a drastic decrease in the TE$_{110}$-mode is the orientation of hole wave functions along the [$\bar{1}$10] direction for the L$_6$ and L$_9$ systems. 

\begin{SCfigure*}
\includegraphics[scale=0.3]{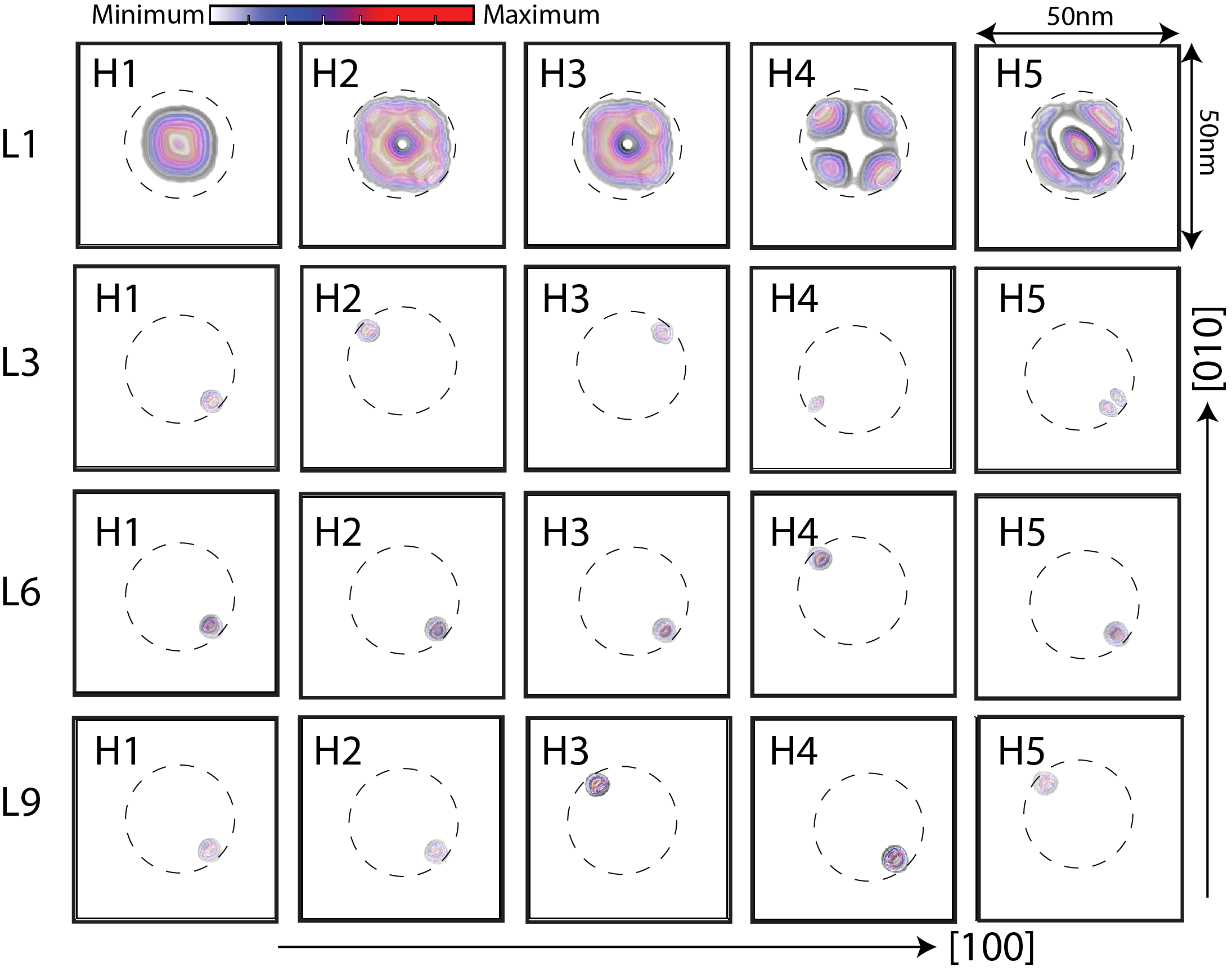}
\caption{Top view of the plots of the highest five valence band states H$_1$, H$_2$, H$_3$, H$_4$, and H$_5$ are shown for L$_1$, L$_3$, L$_6$, and L$_9$ systems. The intensity of the color indicates the magnitude of the hole wave functions: the red color indicating the largest magnitude and the light blue color indicating the smallest magnitude. The dotted circles are marked to guide the eye and indicate the boundary of the QD bases.}
\label{fig:hole_wf}
\end{SCfigure*}

\textbf{\textit{Hole wave functions are oriented along [110] and $[\bar{1}10]$ resulting in $TM_{001} > TE_{110}$}}: Figure ~\ref{fig:hole_wf} plots the top views of the highest five valence band states H$_1$, H$_2$, H$_3$, H$_4$, and H$_5$ for the QD systems L$_1$, L$_3$, L$_6$, and L$_9$. The five hole wave functions for the L$_1$ system have an almost uniform distribution inside the QD region with nearly symmetric shape. Such symmetry will result in approximately equal magnitude of TE-mode along the [110] and $[\bar{1}10]$ directions, as evident in the first row of the Fig. ~\ref{fig:hole_wf}. For the stacks with three, six, and nine QD layers, the distribution of the hole wave functions is oriented along the [110] or [$\bar{1}$10] directions. This [110]/$[\bar{1}10]$-symmetry is mainly due to the strain and piezoelectric potentials that lowers the overall symmetry of the QD system and favours these two directions\cite{Bester_1,Usman_1}. To verify the impact of the strain and piezoelectricity, if we conduct a numerical experiment and switch off their contributions in the electronic structure calculations, the TE$_{\bar{1}10}$/TE$_{110}$ ratio for the L$_6$ system decreases from 10.92 to 0.85 and it decreases from 10.73 to 3.2 for the L$_9$ system. Similar distributions of the hole wave functions are observed for bilayers\cite{Usman_2} and a single QD layer with aspect ratio\cite{Usman_3} (H/B) $\geq$ 0.25. 

The orientation of the hole wave functions determines the magnitude of the TE$_{110}$- and TE$_{\bar{1}10}$-modes since the lowest electron wave function (see Fig. ~\ref{fig:electron_wf}) is symmetrically distributed. For L$_3$, the hole wave functions H$_1$, H$_2$, and H$_5$ are oriented along the [$\bar{1}$10] direction, while the other two hole wave functions H$_3$ and H$_4$ are oriented along the [110] direction. The orthogonal distributions of the hole wave functions in L$_3$ will result in similar [110] and [$\bar{1}$10] TE-modes. The cumulative summations for the TE$_{110}$ and TE$_{\bar{1}10}$-modes arising from these five hole states are indeed nearly equal, with TE$_{\bar{1}10}$ magnitude being slightly larger as can be seen in the Fig.~\ref{fig:intensities}(d).

All of the highest five hole wave functions are oriented along the [$\bar{1}10$] direction in case of the L$_6$ and L$_9$ systems. This results in a strong reduction of the TE$_{110}$-mode. The significant reduction in the TE$_{110}$-mode turns out to be even smaller than the magnitude of TM$_{001}$-mode as can be seen in the Fig.~\ref{fig:intensities}(f, g, j). This change in relative magnitude results in a flip of sign (+ $\rightarrow$ -) for DOP$_{110}$ as indeed measured in the experiment\cite{Inoue_1}. Hence we conclude that the isotropic polarization response demonstrated by the experiment is a result of two factors: (i) increase in the TM$_{001}$-mode due to enhanced HH-LH intermixing and (ii) the reduction of TE$_{110}$-mode due to orientation of holes along the [$\bar{1}10$] direction. 

\vspace{1mm} 

Here we want to point out that the relative magnitude of the TE$_{\bar{1}10}$-mode does not reduce as the size of the QD stack is increased. Even for the largest system under study, L$_9$, the TE$_{\bar{1}10}$-mode is much stronger than the TM$_{001}$-mode. That means if the experimental measurements are performed for DOP$_{\bar{1}10}$, they should still show anisotropy. This is verified by our PL measurements on L$_9$ system shown in Fig. ~\ref{fig:system}(f). These PL measurements on L$_9$ system indicate a positive value for the DOP$_{\bar{1}10}$ (TE$_{\bar{1}10}$ $>$ TM$_{001}$) and a negative value for the DOP$_{110}$ (TE$_{110}$ $<$ TM$_{001}$).                  

\textbf{\textit{Comparison with experimental PL data:}} Table~\ref{tab:table1} summarizes the calculated values of DOP from our model along different directions and compares it with the experimental PL measurements. Here we provide theoretically calculated values of DOP along [100] and [010] directions for comparison purpose because some recent experimental studies have chosen these directions for investigation of the DOP and characterization of the polarization response of the QD systems\cite{Fortunato_1, Saito_1, Ridha_2}.

\begin{table*}
\caption{\label{tab:table1} Comparison of experimentally measured and theoretically calculated DOP for various in-plane TE-mode directions and QD geometry configurations. Column 1: The multi-layer QD system under study. Column 2: The dimensions of the QDs in the stacks: B is base diameter and H is the height of the QD. (V, V) indicates that the QDs are of varying size, base increasing by 1nm and height increasing by 0.25nm as the size of stack increases in the vertical direction. Column 3-6: The values of the DOP calculated from our model. We provide two additional directions for DOP, [100] and [010], for the comparison purpose. Column 7-8: The values of DOP computed from experimentally measured TE- and TM-modes presented in the Fig. ~\ref{fig:system}(b-g) or taken from T. Inoue \textit{et al.}\cite{Inoue_1}}
%\begin{ruledtabular}
%\newcolumntype{M}[1]{>{\raggedright}m{1-8}}
\begin{tabular}{c|c|c|c|c|c|c|c|}
\multicolumn{8}{c}{} \\[3pt]
\cline{1-8}
\multicolumn{2}{|c|}{\textbf{QD Geometry}} & 
\multicolumn{4}{c|}{\textbf{Theoretical Calculations}} & 
\multicolumn{2}{c|}{\textbf{Experiment}}\\[8pt] \cline{1-8}
\multicolumn{1}{|c|}{\textbf{L$_N$}} & \textbf{(B, H) (nm)} & \textbf{DOP$_{100}$} & \textbf{DOP$_{010}$} & \textbf{DOP$_{110}$} & \textbf{DOP$_{\bar{1}10}$} & \textbf{DOP$_{110}$} & \multicolumn{1}{c|}{\textbf{DOP$_{\bar{1}10}$}} \\[6pt] \cline{1-8} 
\multicolumn{1}{|c|}{\multirow{2}{*}{\textbf{L$_1$}}} & (20, 3.5) & 0.995 & 0.996 & 0.996 & 0.996 
& \multirow{2}{*}{0.7} & \multirow{2}{*}{0.71} \\[5pt] \cline{2-6}
\multicolumn{1}{|c|}{} & (20, 4.0) & 0.999 & 0.999 & 0.999 & 0.999 & & \\[5pt] \cline{1-8}
\multicolumn{1}{|c|}{\multirow{3}{*}{\textbf{L$_3$}}} & (20, 3.5) & 0.922 & 0.884 & 0.9046 & 0.933 
& \multirow{3}{*}{0.67\cite{Inoue_1}} & \multirow{2}{*}{} \\[5pt] \cline{2-6}
\multicolumn{1}{|c|}{} & (20, 4.0) & 0.77 & 0.77 & 0.833 & 0.836 & & \\[5pt] \cline{2-6}
\multicolumn{1}{|c|}{} & (V, V) & 0.929 & 0.929 & 0.916 & 0.94 & & \\[5pt] \cline{1-8}
\multicolumn{1}{|c|}{\multirow{3}{*}{\textbf{L$_6$}}} & (20, 3.5) & 0.594 & 0.594 & 0 & 0.747 
& \multirow{3}{*}{0.46\cite{Inoue_1}} & \multirow{2}{*}{} \\[5pt] \cline{2-6}
\multicolumn{1}{|c|}{} & (20, 4.0) & 0.548 & 0.46 & -0.244 & 0.72 & & \\[5pt] \cline{2-6}
\multicolumn{1}{|c|}{} & (V, V) & 0.733 & 0.733 & 0.73 & 0.76 & & \\[5pt] \cline{1-8}
\multicolumn{1}{|c|}{\multirow{3}{*}{\textbf{L$_9$}}} & (20, 3.5) & 0.6 & 0.6 & 0.445 & 0.652 
& \multirow{3}{*}{-0.6\cite{Inoue_1}, -0.36} & \multirow{3}{*}{0.66} \\[5pt] \cline{2-6}
\multicolumn{1}{|c|}{} & (20, 4.0) & 0.38 & 0.371 & -0.45 & 0.603 & & \\[5pt] \cline{2-6}
\multicolumn{1}{|c|}{} & (20, 4.5) & 0.45 & 0.446 & -0.47 & 0.43 & & \\[5pt] \cline{1-8}
\end{tabular} 
%\end{ruledtabular}
\end{table*}

The hole state symmetries shown in the Fig. ~\ref{fig:hole_wf} show strong alignment in the [110] and [$\bar{1}10$] directions indicating that symmetries to the [100] and [010] directions will be almost equivalent. Therefore the values of the DOP are nearly equal for the [100] and [010] directions as mentioned in the table ~\ref{tab:table1}. The PL measurements along these two directions will not exhibit isotropic polarization even for the L$_9$ system. Similar results were found in an earlier experimental study by P. Ridha \textit{et al.}\cite{Ridha_2}. In their study of the polarization properties of multi-layer stacks based on TE$_{010}$-mode, they conclude that such systems can not provide isotropic polarization and columnar QDs are the only choice.    

For the system L$_1$ containing only a single QD layer, our calculated DOP values in all the directions are very close to 1.0, exhibiting a very strong polarization anisotropy (TE-mode $\gg$ TM-mode). The experimentally measured values for this system are relatively low ($\sim$0.7), but they also show that the value of the DOP remains nearly the same irrespective of the measurement direction. While the reason for difference between our calculated value and the measured value for this QD system is not very clear, other theoretical studies using k.p method\cite{Saito_1, Sheng_1} on single QD layers with similar dimensions have also presented the values of the DOP close to 1.0.

\textbf{\textit{Sensitivity of the DOP with the QD stack geometry parameters:}} Since the heights of the QDs in the experimental TEM images are not very clear (see ~\ref{fig:system}(a)), so we simulate various geometry configurations of the QD stacks and provide the values of the DOP in the table \ref{tab:table1}. This data serve as a measure of the sensitivity of the DOP with respect to the QD stack geometry parameters and provide a guide to experimentalists to explore the design space of such complex multi-million atom systems. From the table ~\ref{tab:table1}, as the QD stack height increases in the L$_1$ $\rightarrow$ L$_3$ $\rightarrow$ L$_6$ $\rightarrow$ L$_9$, the value of the degree of polarization reduces. The reduction in the value of the DOP is larger for the stacks with H=4.0nm as compared to the stacks with H=3.5nm. This is due to the fact that the larger height of QDs in the stack results in stronger coupling between the QDs. This implies a stronger HH-LH intermixing resulting in larger magnitude of the TM$_{001}$-mode. 

The dependence of the DOP on the height (H) of the QDs inside the stacks is an unknown factor. The calculated values of the DOP in the table ~\ref{tab:table1} shows that the DOP becomes very sensitive to the height of the QDs inside the stack as the size of the stack grows larger. For the systems L$_1$ and L$_3$, the increase in the height (H) from 3.5nm to 4.0nm results in a small decrease in the values of the DOP. However, for a same change in the value of H, the DOP$_{110}$ significantly decreases from 0 to -0.244 and from 0.445 to -0.45 for the L$_6$ and L$_9$ systems, respectively. This implies that an isotropic polarization response (DOP $\sim$ 0) can either be achieved from the L$_6$ stack with H=3.5nm, or from a stack with fewer number of QD layers and H=4.0nm. We therefore propose that the polarization response of the QD stacks can be tuned by not only increasing the number of QD layers (a parameter tuned in the past experimental studies\cite{Inoue_1, Ridha_1, Ridha_2, Saito_1, Kita_1}), but also by controlling the height (H) of the individual QDs inside the stacks.   

The calculations on the L$_3$ and L$_6$ systems, where the size of QDs is increased as the stack size increases, indicate that such stacks with non-identical QD layers will exhibit relatively higher values of the DOP and hence will not be suitable for isotropic polarization response. The same is true for the case when the height of the QDs in the L$_9$ system is increased to 4.5nm such that the adjacent QD layers touch each other (approaches columnar QD limit). Based on the comparison of the calculated and the measured values of the DOP for the L$_9$ system in the table ~\ref{tab:table1}, we estimate that the dimensions of the QDs inside the stacks are approximately B=20nm and H=4nm. 

\textbf{\textit{Conclusions:}} This article presents a detailed analysis of the polarization response of the multi-layer QD stacks by PL measurements and through a set of systematic multi-million atom tight binding electronic structure calculations. Our theoretical results follow the trends of the experimental measurements on quantum dot stacks containing single, three, six and nine QD layers and provide significant physical insight of the complex physics involved by analysing the strain profiles, the band edge diagrams, and the wave function plots. The experimentally measured PL data for the nine quantum dot stack reveals a unique property by indicating a significant difference in the DOP for the [110] and [$\bar{1}10$] directions. We explain here that this difference is due to the orientation of the hole wave functions along the [$\bar{1}10$] direction that results in significant reduction of the TE-mode along the [110] direction. We suggest that the isotropic polarization response from the multi-layer QD stacks is due to two factors: (i) the reduction of the TE$_{110}$-mode direction and (ii) the increase in the TM$_{001}$-mode due to enhanced LH-HH intermixing. Our result presented in this paper for various geometry configurations serve as guidance for the experimentalists to design future QD based optical devices. A flip of sign for the DOP in our PL measurements and theoretical calculations as the size of QD stack increases indicates significant potential to achieve polarization insensitive response from multi-layer QD stack systems.               
	 
\textbf{\textit{Acknowledgements:}} Muhammad Usman gratefully thanks Dr. Susannah Heck (OCLARO, UK) for useful discussions about the polarization response of the QD samples. Computational resources accessed through the nanoHUB.org application "Workspace" operated by the NSF funded Network for Computational Nanotechnology are acknowledged.  The back-end supercomputers are operated by the Rosen Center for Advanced Computing (RCAC), Purdue University.

%----------------------------------------------------------------------
% References
%----------------------------------------------------------------------

\end{document}